\documentclass[preprint]{aastex}
\usepackage{natbib}
\begin{document}

\title{The First Circumstellar Disk Imaged in Silhouette with Adaptive Optics: MagAO Imaging of Orion 218-354}
\author{Katherine B. Follette\altaffilmark{1}, Laird M. Close\altaffilmark{1}, Jared R. Males\altaffilmark{1,4}, Derek Kopon\altaffilmark{2}, Ya-Lin Wu\altaffilmark{1}, Katie M. Morzinski\altaffilmark{1,4}, Philip Hinz.\altaffilmark{1}, Timothy J. Rodigas\altaffilmark{1}, Alfio Puglisi\altaffilmark{3}, Simone Esposito\altaffilmark{3}, Armando Riccardi\altaffilmark{3}, Enrico Pinna\altaffilmark{3}, Marco Xompero\altaffilmark{3}, Runa Briguglio\altaffilmark{3}}
\altaffiltext {1}{Steward Observatory, The University of Arizona, 933 N Cherry Ave, Tucson, AZ 85721, USA}
\altaffiltext {2}{Max Planck Institute for Astronomy, K\"{o}nigstuhl 17, 69117 Heidelberg, Germany}
\altaffiltext {3}{INAF - Osservatorio Astrofisico di Arcetri, Largo E. Fermi 5, I-50125, Firenze, Italy}
\altaffiltext {4}{NASA Sagan Fellow}

\begin{abstract}
We present high resolution adaptive optics (AO) corrected images of the silhouette disk Orion 218-354 taken with Magellan AO (MagAO) and its visible light camera, VisAO, in simultaneous differential imaging (SDI) mode at H$\alpha$.  This is the first image of a circumstellar disk seen in silhouette with adaptive optics and is among the first visible light adaptive optics results in the literature. We derive the disk extent, geometry, intensity and extinction profiles and find, in contrast with previous work, that the disk is likely optically-thin at H$\alpha$. Our data provide an estimate of the column density in primitive, ISM-like grains as a function of radius in the disk. We estimate that only $\sim$10\% of the total sub-mm derived disk mass lies in primitive, unprocessed grains. We use our data, Monte Carlo radiative transfer modeling and previous results from the literature to make the first self-consistent multiwavelength model of Orion 218-354. We find that we are able to reproduce the  1-1000$\mu$m SED with a $\sim$2-540AU disk of the size, geometry, small vs. large grain proportion and radial mass profile indicated by our data. This inner radius is a factor of $\sim$15 larger than the sublimation radius of the disk, suggesting that it is likely cleared in the very interior. 
\end{abstract}

\section{Introduction}

Silhouette disks were first discovered in 1994 by \citet{Odell:1994}. A single dark silhouette, Orion 183-405,  was seen against the bright H$\alpha$ emission of the Orion Nebula in their HST Wide Field Camera (WFC) images of the region.  \citet{Odell:1996} followed this with a Wide Field Planetary Camera 2 (WFPC2) HST survey of the brightest regions of the nebula and discovered six additional disks.  To this day, images of the Orion silhouette disks provide some of the most conclusive evidence for the existence and sizes of dusty circumstellar disks around young stars. 

 \citet{McCaughrean:1996} conducted the first detailed analysis and modeling of silhouette disks. The primary conclusion of their modeling of the silhouettes identified in \citet{Odell:1996} was that all were best-fit by optically-thick opaque disk models with exponential edges. As column density cannot be determined from an optically-thick disk profile, they were only able to place lower limits on the amount of material in these disks. For the disk that is the focus of this work, Orion 218-354, their best-fit models suggests an r=0$\farcs$54 disk with an inclination of 60$^{\circ}$ and a total mass of 2.4$\times$10$^{-5}$M$_{\sun}$.
 
Later surveys of the Orion region with WFPC2 \citep{Bally:2000} and ACS \citep{Ricci:2008} revealed many more of these disks, bringing the total to 28 known Orion silhouettes. Some have been followed up at other wavelengths, including the sub-mm \citep[e.g.][]{Mann:2010}, thermal infrared \citep[e.g.][]{Hayward:1997} and x-ray \citep[e.g.][]{Kastner:2005}, however except for the very largest silhouette disk (Orion 114-426, r$>$1"), they have not been imaged from the ground in silhouette until now.  
 
For many years following the pioneering HST observations, ground-based imaging of these disks in silhouette were precluded by their small size (r$\le$1``), requiring higher resolutions than were available with seeing limited images. Adaptive optics has long been capable of delivering such resolutions in the infrared, however the bright nebular emission lines where these disks appear in silhouette all lie in the optical regime, blueward of the operating wavelengths of most AO systems. Is is only with the high actuator pitch of modern adaptive secondary mirrors that correction on the necessary spatial scales for observations at optical wavelengths has been achieved. 

The central stars of the silhouette disks are also relatively faint (R$\ge$11), putting them outside of the working range of most AO wavefront sensors. Pyramid wavefront sensors, however, allow for binning and can achieve correction on fainter stars.  MagAO is among the first modern AO systems with the ability to achieve the necessary resolutions for such imaging from the ground, among the first systems with the ability to lock on sufficiently faint natural guide stars, and the first large telescope with an AO-optimized visible light camera capable of imaging at the necessary wavelengths to see disks in silhouette \citep{Kopon:2009, Kopon:2012, Close:2012, Males:2012, Follette:2010}. 

\section{Observations and Data Reduction}

Observations of Orion 218-354 were conducted on December 6, 2012 as part of the commissioning of Magellan's Adaptive Secondary AO System (MagAO).   MagAO is a natural guide star (NGS) facility instrument of the 6.5m Magellan Clay Telescope at Las Campanas Observatory. The commissioning performance of the system is detailed in Close et al. 2013, (ApJ, accepted). 

The complete data set consists of 72 30 second images of Orion 218-354, however only the best 38 images were used, for a total integration time of 19 minutes.  The 1024x1024 pixel CCD47 VisAO camera was used in simultaneous differential imaging (SDI) mode in which a Wollaston prism separates the VisAO beam into two beams of approximately equal brightness (for unpolarized sources). These beams are each passed through a separate narrowband filter, one centered on the spectral line of interest (H$\alpha$, [OI] or [SII]) and one on the neighboring continuum. For Orion 218-354, the H$\alpha$ SDI filter set was used, with one filter of width 4.6nm centered on H$\alpha$ at 656.6nm and one of width 6.1nm centered on the continuum at 642.8nm (hereafter ``continuum"). 

Two different image rotator orientations were used, and images were obtained at three different chip locations for each rotator angle in order to distinguish filter artifacts from background structure in the nebula. To create the final images, the raw images for both channels were dark subtracted, flat fielded, rotated to a common orientation, and registered  

The median combinations of the H$\alpha$ and continuum channels are shown in Figure 1 in the top left and bottom left panels. Background structure in the nebula is apparent at  H$\alpha$, and is lacking in the continuum channel, as expected. A secondary star lies 2" to the SE of Orion 218-354 in both, and the dark silhouette of the disk is visible in H$\alpha$ even before PSF subtraction. 

Due to the guide star brightness of R=12.5, requiring wavefront sensor binning to 2x2, even though significant improvement in image FWHM was obtained (FWHM$\sim$.1" vs. 0.7-1.2" seeing), Strehl ratio was still very low.  The PSF therefore consists of a single profile, rather than the typical core+halo profile seen with higher Strehls. Despite the handicaps of short wavelength and dim guide star, the PSF of the central star is exceptionally well measured by the simultaneous acquisition of the continuum channel, and can be robustly removed. 

PSF subtraction in MagAO's SDI mode is simpler than ``traditional" PSF subtraction because the continuum channel is a simultaneous probe of the PSF at a nearly identical wavelength, and PSF structure is therefore identical. The only non-common optics between the channels are their respective SDI filters, and filter artifacts are easily removed by taking the median over several rotation angles.  Because most stellar spectra have real structure (absorption) at H$\alpha$ that varies with spectral type, there is no absolute scaling between the filters for subtraction. Instead, it is determined on a case-by-case basis from the ratio of the peaks in both the H$\alpha$ and continuum images.

The dominant source of error in the scaling of the PSF for subtraction comes from the 45$^{^\circ}$ tertiary mirror that feeds MagAO, which is made of freshly coated (2012) aluminum with a reflectivity of $\sim$97\%. Because the $\sim$3\% of light absorbed by the coating may have a preferential polarization, and the orientation of the Wollaston relative to the tertiary changes as the instrument rotates, the scaling between the channels may change by as much as 3\% over the course of observations taken with the rotator on. 

In the case of Orion 218-354, the continuum image was scaled by a factor of 1.03 before subtraction from the H$\alpha$ image based on the ratio of the peaks. Both Orion 218-354 and the secondary star in the images are late-type stars \citep{Hillenbrand:1997, Terada:2012} with very little spectral structure in the H$\alpha$ region, which allowed us to use aperture photometry of the secondary star to verify this scaling. We found an identical 1.03 scaling based on this methodology. 

The resulting PSF subtracted image is shown in the bottom righthand panel of Figure 1. The error in this subtraction is likely much less than 3\% because of the short duration of the observations, however errors derived from a full 3\% different scaling of the PSF are shown throughout this letter, and are shown in Figure 2.

The upper righthand panel of Figure 1 shows the HST ACS image of Orion 218-354 from \citet{Ricci:2008} taken in the F658N (H$\alpha$) filter. The scale of the image is the same as the MagAO images to allow for direct comparison of spatial structure. The secondary star and the background structures in the nebula appear in both the MagAO and HST H$\alpha$ images. The primary difference is that the central star in the HST image is heavily saturated, making it difficult to recover information about the inner disk 

\section{Results and Analysis}

Light blue isophotal contours are overplotted on the continuum-subtracted H$\alpha$ image in Figure 1.  They reveal that the brightness of Orion 218-354 at H$\alpha$ falls steadily towards the center of the disk, calling into question it's classification as optically-thick. The radial profile along the major and minor axes is shown in Figure 2. The shape of these profiles remains the same for a range of scaled PSF subtractions. The profiles were binned to 7 pixels ($\sim$0$\farcs$055, half of the measured FWHM of the continuum PSF). The diffraction limit of the MagAO system at H$\alpha$ is $\sim$20mas,  and the system routinely achieves 6 and 7 pixel FWHM at this wavelength.  This drop in performance was due to (a) relatively poor seeing for the site, ranging from 0$\farcs$7 at the start of the observations to 1$\farcs$2 at the finish, and (b) the faintness of the guide star, requiring binning of the pyramid pupils to 2x2. 

Also overplotted on the radial profile in Figure 2 is the optically-thick best-fit disk model of \citet{McCaughrean:1996} (a 0$\farcs$54 60$^{\circ}$ inclined opaque disk seen against the nebular background) convolved with the Magellan continuum PSF. Several sizes of PSF were used for convolution to investigate the effect of the PSF wings. The PSF reaches the nebular background level at r$\sim$75 pixels, therefore we feel that the 150x150 pixel PSF convolution shown in purple in Figure 2 is the most robust. Enlarging the PSF has a pedestal effect on the convolved profile, effectively ``filling in" the central part of the disk with nebular background light, however the shape and character of the profile remain nearly identical, as shown by the yellow 250x250pixel PSF convolution profile (normalized to match the 150x150 profile at the inner and outer edge of the disk). Both convolved model profiles are immediately apparent as entirely different in shape and character from the observed radial intensity profile. 

The MagAO profile is well fit by a powerlaw with exponent 0.46 and an exponential cutoff beyond $\sim$0$\farcs$75 ($\sim$300AU), shown overplotted in red. The flux in the MagAO radial profile reaches the background level at r=1$\farcs$3, which corresponds to a disk radius of $\sim$540$\pm$60 AU at 414pc \citep{Menten:2007}, significantly larger than the \citet{McCaughrean:1996} value of 0$\farcs$54.  Our best-fit to the outer disk suggest an ellipticity of 0.72$\pm$0.08, corresponding to a disk inclination of 44$\pm$5$^{\circ}$ (slightly smaller than the \citet{McCaughrean:1996} value of 60$^{\circ}$), and a major axis PA=72$\pm$10$^{\circ}$ East of North. 

Intensity in the difference image was translated to extinction by:
\begin{displaymath}
A_{H\alpha}=-2.5\log(I/I_{B})
\end{displaymath}
where I is the intensity in each pixel and I$_{B}$ is the background intensity in the nebula. We converted this extinction value at H$\alpha$ (A$_{H\alpha}$) to an extinction value at V (A$_{V}$) by linear interpolation of extinctions at bracketing wavelengths per \citet{Mathis:1990}, resulting in a factor of 1.22 increase in the extinction value from H$\alpha$ to V band.  

We used the empirical relationship between A$_{V}$ and hydrogen column density (N$_{H}$) derived by \citet{Bohlin:1978} (N$_{H}$=1.87$\times$10$^{21}$$\times$A$_{V}$ cm$^{-2}$) to convert A$_{V}$ to N$_{H}$. We multiplied this value for N$_{H}$ by the physical size of a MagAO pixel at 414pc, and by the mass of a hydrogen atom to get a disk dust mass estimate of 2.3$\pm$1$\times$10$^{-5}$M$_{\sun}$, or approximately 7.5M$_{\earth}$. Errors were estimated from an equivalent conversion of the difference image with a 3\% different PSF scale factor. 

It is important to note that the A$_{V}$ to N$_{H}$  conversion value that we've employed is appropriate only for an extinction curve slope R$_{V}$=3.1, which corresponds to the diffuse interstellar medium. Larger grains have poorer reddening efficiency and are therefore ``missing" in this approximation. In other words, our mass estimate probes only primitive ISM-like grains in the disk. 

Assuming a 100:1 gas:dust mass ratio, we convert this value to a total disk mass of $\sim$2.3$\pm$1$\times$10$^{-3}$M$_{\sun}$. As a probe of the total disk mass, including large grains, the sub-mm disk mass estimate of 0.0237M$_{\sun}$  \citep{Mann:2010} is much more robust. The order of magnitude difference between our mass estimate and the sub-mm estimate suggests that just $\sim$10\% of the disk mass lies in primitive grains that absorb efficiently at H$\alpha$. 

We believe this value to be robust for several reasons. First, any foreground H$\alpha$ emission should be relatively uniform across the disk, and will not contribute to the differential extinction measurement. Secondly, scattered light from grey grains in the disk should be virtually identical between the two SDI filters, and therefore should be removed by PSF subtraction. Finally, because the radial profile of our best PSF subtraction approaches zero in the interior, there is little room to achieve a higher integrated extinction.  

This small grain dust mass, as well as the shape of the radial intensity profile shown in Figure 2, are suggestive of a disk that is optically-thin at H$\alpha$ for r$\gtrsim$25AU (our innermost resolution element), a surprising result given previous work on silhouette disks. This low optical depth may be due to any number of factors, including small grain blowout, blowout due to ionizing radiation from the Trapezium, or grain growth in the disk. 

An extinction/column density profile derived following the same procedure described above is shown in Figure 3 in both linear and log (inset) space. The inner disk is well fit by a powerlaw with exponent -1.43. An exponential disk edge reproduces the deviation from this powerlaw at large radii, as shown by the overplotted best-fit intensity profile. The primary source of potential error in this and other measurements of the observed radial profile is that they have been effectively convolved with the instrumental PSF and may deviate somewhat from the "true`` disk profile. We leave investigation of this effect for future work. 

This extinction profile is a probe of the integrated mass density profile of ISM grains in the disk. Total mass in a given column scales with both the density at the midplane and the scale height of the disk at that radius (M(r) $\propto$ Z$\times\rho_{midplane} \propto$ r$^{\beta}\times$r$^{-\alpha}$). Assuming a geometrically flat disk where scale height Z $\propto$ r$^{1}$ ($\beta$=1), the extinction powerlaw gives us an estimate of the midplane density profile:$\rho_{midplane}\propto$ r$^{-2.43\pm0.3}$ ($\alpha$=2.43$\pm$0.3). This is well within the range of values commonly assumed for the midplane density distribution.  If the disk is flared ($\beta$$>$1), this is instead a lower limit ($\alpha$$>$2.43$\pm$0.3).

To support our conclusions about the geometry and mass distribution in the disk, we've used our derived parameters and a series of simple assumptions as inputs to the Whitney 3D Monte Carlo Radiative Transfer Code \citep[][Whitney et al. 2013, ApJS, accepted]{Whitney:2003,Whitney:2003a} to generate a 0.1-1000$\mu$m SED. Input parameters are given in Table 1, and the generated SED is shown in Figure 4. 

Photometric points from the literature are overplotted on the SED. They show that emission in Orion 218-354 is at or only slightly in excess of photospheric emission for all wavelengths L` and shortward. Only the sub-mm photometry (and, of course, the observed silhouette) strongly suggest the presence of a circumstellar disk. 

The SED also suggests the presence of a small inner gap in the disk, as models that extend inward to the sublimation radius drastically overproduce NIR flux. A model with $r_{disk, in}$=$r_{sub}$ is shown overplotted as a dashed line on Figure 4 to demonstrate this. In order to reproduce the L' photometry of \citet{Muench:2002} and \citet{Terada:2012a}, the disk cannot extend farther inward than $\sim$15r$_{sub}$, or about 2AU. Therefore, the innermost 2AU of the Orion 218-354 disk is likely cleared of material. 

\section{Conclusion}
In this study we have presented the first ground-based adaptive optics images of a circumstellar disk seen in silhouette. We derived geometric parameters of Orion 218-354 (r$\sim$540AU, $i\sim$46$^{\circ}$, PA$\sim$72$^{\circ}$) that suggest it is more extended and less inclined than previous observations would suggest. The radial intensity profile of our data is inconsistent with the opaque, optically-thick disk suggested by earlier modelers, as it shows a steady increase with radius in the amount of background nebular flux that passes through the disk. This suggests that the column density in small grains decreases steadily with radius. The fact that the amount of absorption does not plateau at small radii, as convolution of optically-thick disk models with our observed PSF suggest it would have, suggests that the disk is optically-thin at H$\alpha$, at least as far inward as our innermost resolution element.  Using the extinction/column density profile, we estimate the mass as a function of radius at the midplane in the disk assuming that it is geometrically flat, and find that it is proportional to r$^{-2.43}$. 

The integrated column density of the disk translates to a mass of $\sim$2.3$\pm$1$\times$10$^{-5}$M$_{\sun}$ in primitive ISM-like grains, and a total disk mass of $\sim$2.3$\pm$1$\times$10$^{-3}$M$_{\sun}$ (assuming a 100:1 gas to dust mass ratio). The 880$\mu$m sub-mm continuum flux for this disk, on the other hand, suggests a disk mass that is higher by an order of magnitude (0.0237M$_{\sun}$). This suggests that the majority of the mass in this disk (90\%) may lie in grains that have grown beyond ISM-like sizes. 

We model the 1-1000$\mu$m SED of Orion 218-354, showing that the sub-mm mass estimate and our derived parameters are consistent with published multiwavelength photometry of the disk. The NIR region of the SED suggests that the inner disk contains a small gap, which we estimate at 15r${sub}$ or $\sim$2AU.  

This study demonstrates the power of modern adaptive secondary AO systems to achieve atmospheric correction into the visible wavelength regime, the ability of pyramid wavefront sensors to achieve excellent correction on faint guide stars, and the power of simultaneous differential imaging. Due to these complimentary technologies, the future of circumstellar disk imaging from the ground at visible wavelengths is bright.

\begin{acknowledgements}
We would like to thank Barb Whitney, Joan Najita and Kate Su for their insights regarding these observations. 
\end{acknowledgements}

\clearpage

\begin{table} [ht]
\centering
\begin{tabular} {c c}
\hline\hline
Parameter & Model Input(s)\\
\hline
M$_{star}$ & 1.9M$_{\sun}$\\
R$_{star}$ & 2.4R$_{\sun}$\\
T$_{star}$ & 5272K\\
A$_{V}$ & 1.51\\
M$_{disk}$ & 0.02373M$_{\sun}$\\
$r_{disk, in}$ & $15r_{sub}$\\
$r_{disk, out}$ & 538AU\\
$i$ & 46$^{\circ}$\\
f$_{d}$ & 0.9\\
Z$_{100}$ & 7.5AU \\
$\alpha$ & 2.43\\
$\beta$ & 1.0\\
\hline\hline
\end{tabular}
\label{table:inputs}
\caption{best-fit Model Input Parameters.  f$_{d}$ represents the fraction of the disk's dust mass in large grains, Z$_{100}$ is the scale height at 100 AU, $\alpha$ is the exponent for the radial midplane density, and $\beta$ is the exponent for the radial scale height. The stellar and extinction parameters are from \citet{Hillenbrand:1997}; the disk mass is from \citet{Mann:2010}. All other parameters were derived from our observations as described in the text with the exception of Z$_{100}$, for which the model was found to be insensitive to a range of reasonable parameters (values 5-10AU),  and $r_{disk, in}$, which was determined iteratively from the fit to existing J-L` photometry. The large grain dust prescription is Model 1 of \citealt{Wood:2002}, a mixture of amorphous carbon and astronomical silicates with a maximum grain size of 1000$\mu$m. The small grain dust prescription is the ISM grain model of \citealt{Kim:1994}, a mixture of silicate and graphite with a maximum grain size of 0.28$\mu$m.}
\end{table}

\begin{figure}
\includegraphics[scale=0.8]{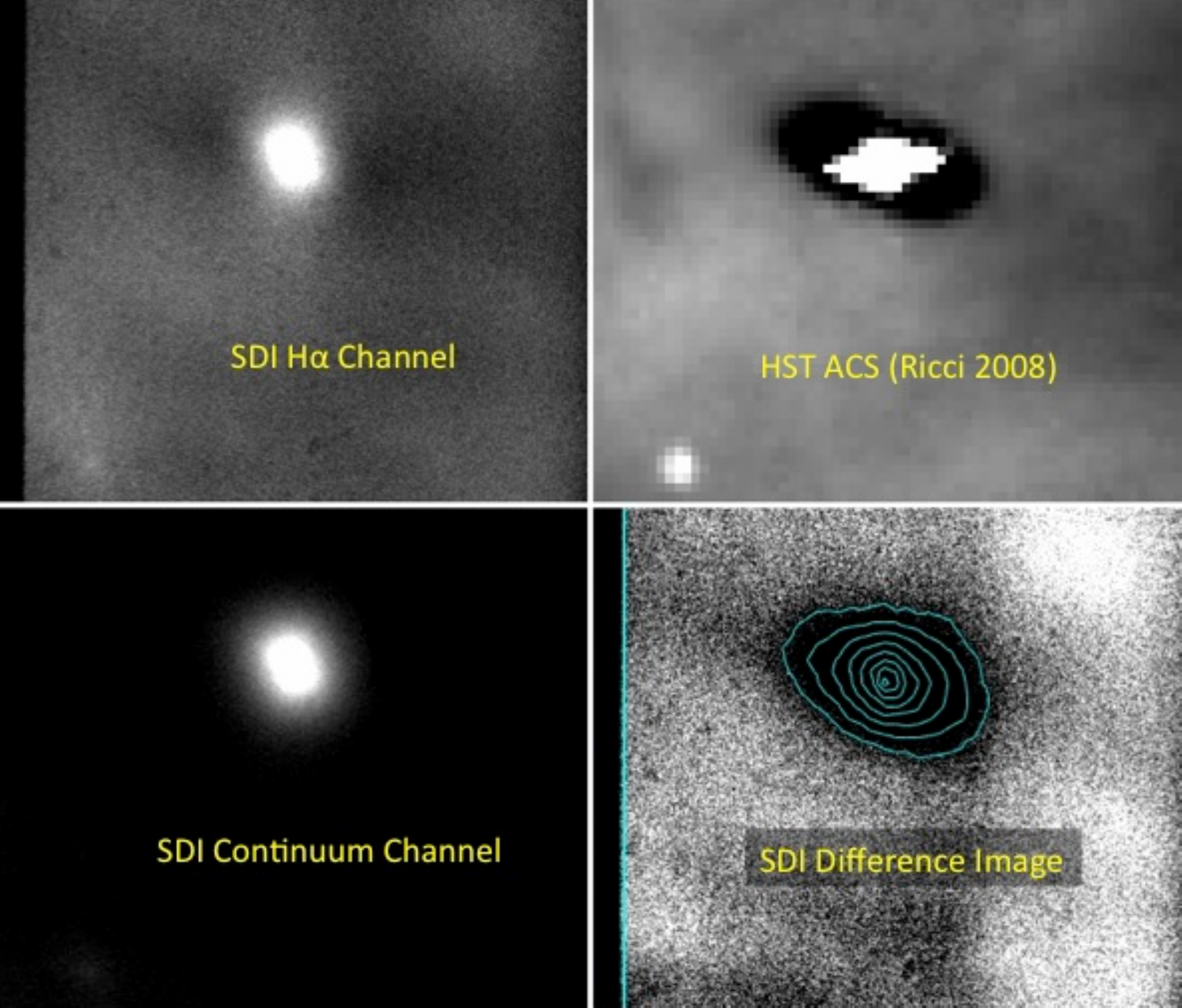}
\caption{Upper left: MagAO H$\alpha$ channel. Lower left: MagAO 643nm continuum channel. Lower right: MagAO difference image created by subtracting the scaled continuum (lower left) from H$\alpha$ (upper left). Upper right: HST ACS image of the same disk \citep{Ricci:2008}. All images are North up, East left and have the same physical scale. Note that the disk is visible in all panels except the continuum image, where it shouldn't be visible. Subtraction of the simultaneous PSF provided by the continuum channel effectively removes both the primary star and a secondary star 2" to the SE, isolating H$\alpha$ emission from the background nebula. The disk stands out starkly in silhouette in this (lower right) image, and light blue contours reveal that the background flux from the nebula is attenuated to an increasing degree as column density increases towards the center of this disk. The extent of the contours all the way to the central star is suggestive of an optically-thin disk at H$\alpha$.  \label{fig1}}
\end{figure}

\begin{figure}
\includegraphics[scale=1]{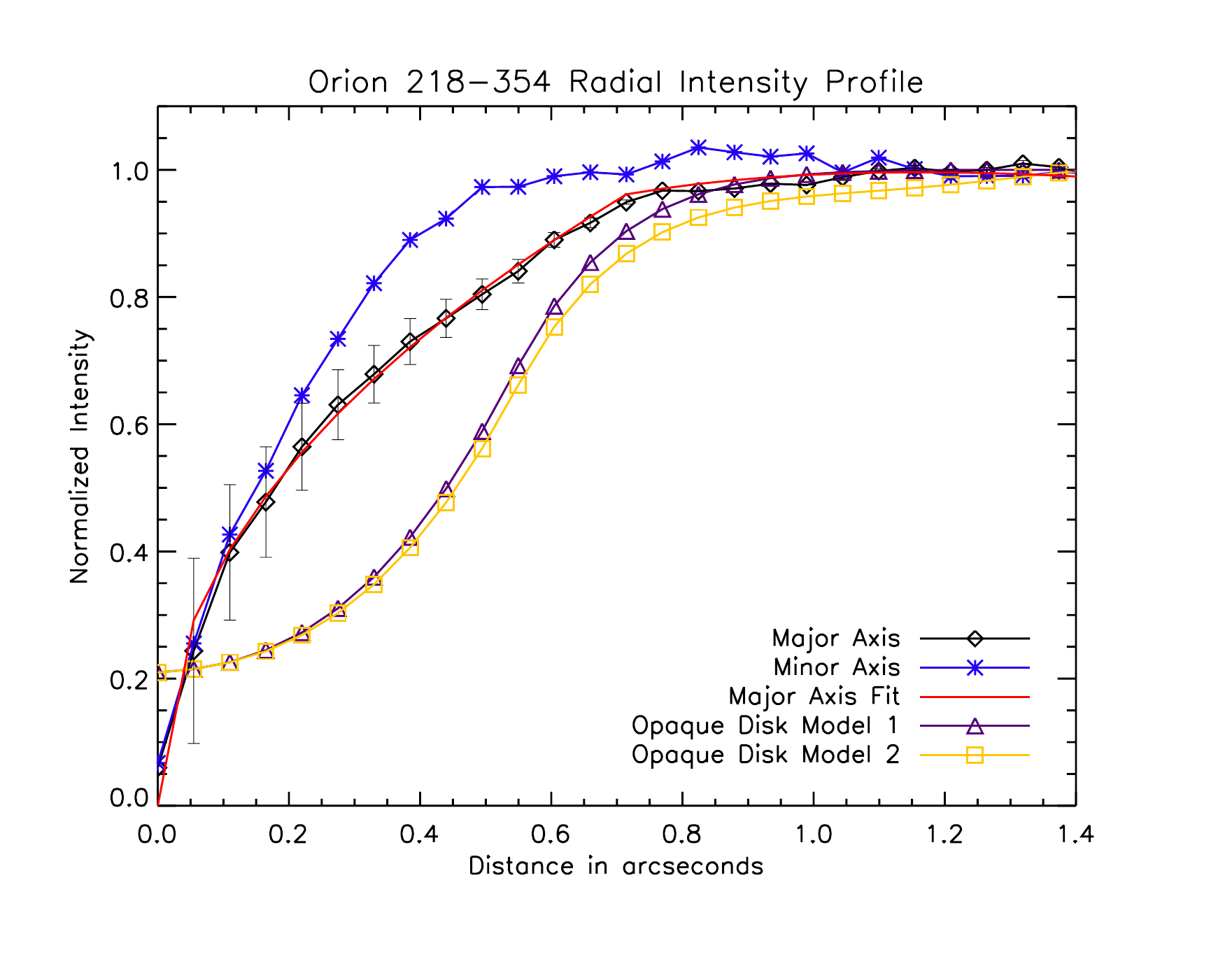}
\caption{Radial intensity profile of the Orion 218-354 disk normalized to the background nebular intensity level. Profiles are shown for the major (black diamonds) and minor (blue asterisks) axes. An opaque optically-thick disk model with the best-fit parameters derived in \citet{McCaughrean:1996} convolved with both 150x150 and 250x250 MagAO continuum PSFs are shown overplotted (purple triangles, yellow squares). In order to demonstrate that the character of the profiles is the same, the 250x250 pixel model has been normalized to the 150x150 model to remove the pedestal effect caused by ``filling in" the inner disk with nebular light. The inconsistency in shape and character of these models with our data is striking, and the steadily decreasing profile we observe suggests an optically-thin disk. The performance of a modern AO system, as well as the simpler methodology of SDI-mode PSF subtraction, gives us a clear advantage over the HST data, in which the central star is heavily saturated.  The best-fit to the MagAO profile, a powerlaw with exponent 0.46 and an exponential cutoff beyond $\sim$0$\farcs$75, is shown overplotted in red. Errors due to a 3\% deviant scaling of the continuum PSF for subtraction are also shown. \label{fig2}}
\end{figure}

\begin{figure}
\includegraphics[scale=1]{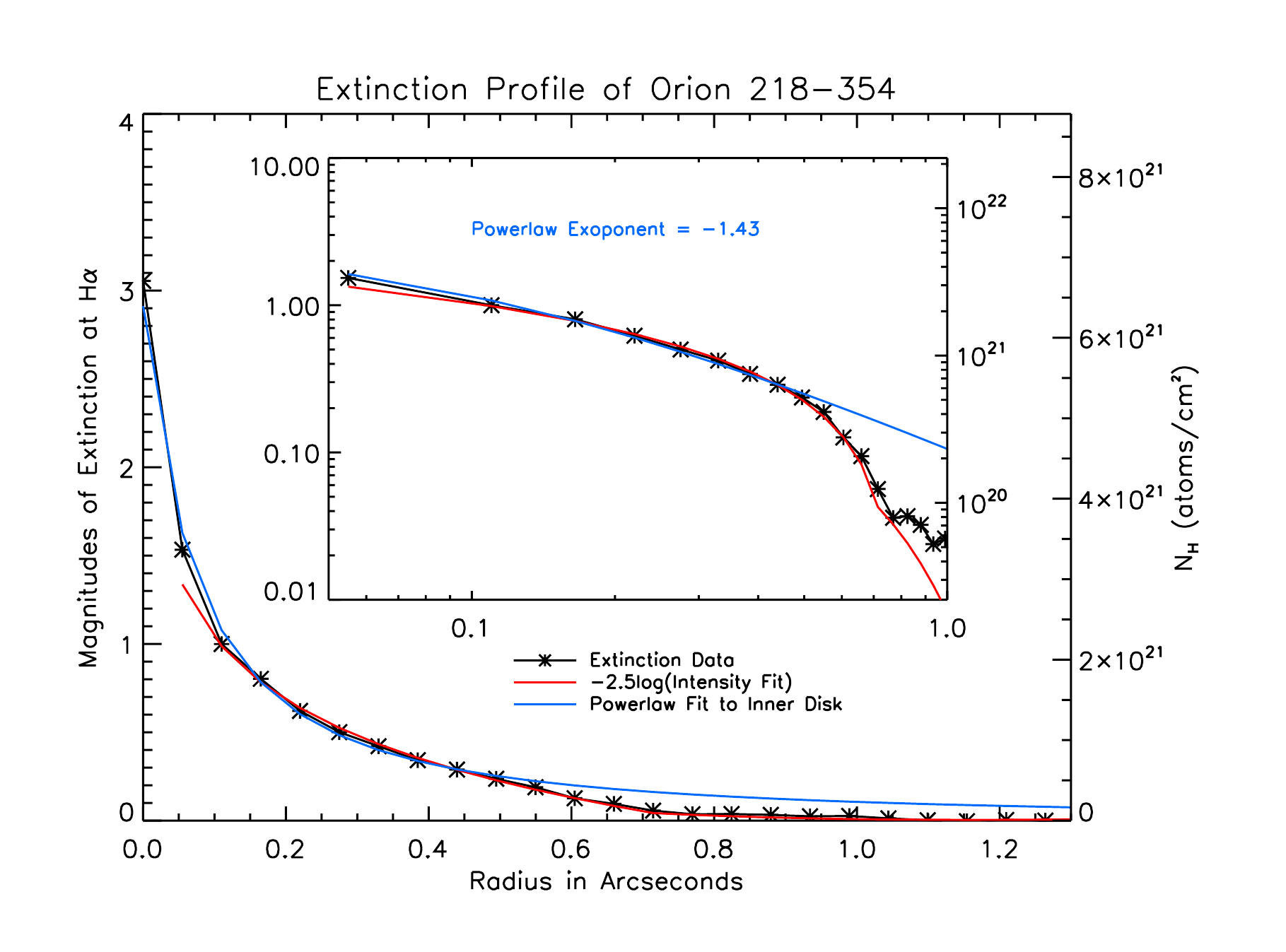}
\caption{Radial extinction/column density profile of the Orion 218-354 disk in linear and log (inset) space. Overplotted on the data (black asterisks) in both plots are the Figure 2 intensity fit translated to units of extinction (red line) and a best-fit powerlaw to the extinction profile in the inner disk (blue line), which has an exponent of -1.43. This value is used to estimate the radial mass distribution of the disk ($\rho \propto $r$^{-\alpha}$), under assumptions described in detail in the text. This extinction should be considered to represent the column density of primitive, ISM-like material in the disk, and not of the disk as whole. \label{fig3}}
\end{figure}

\begin{figure}
\includegraphics[scale=1]{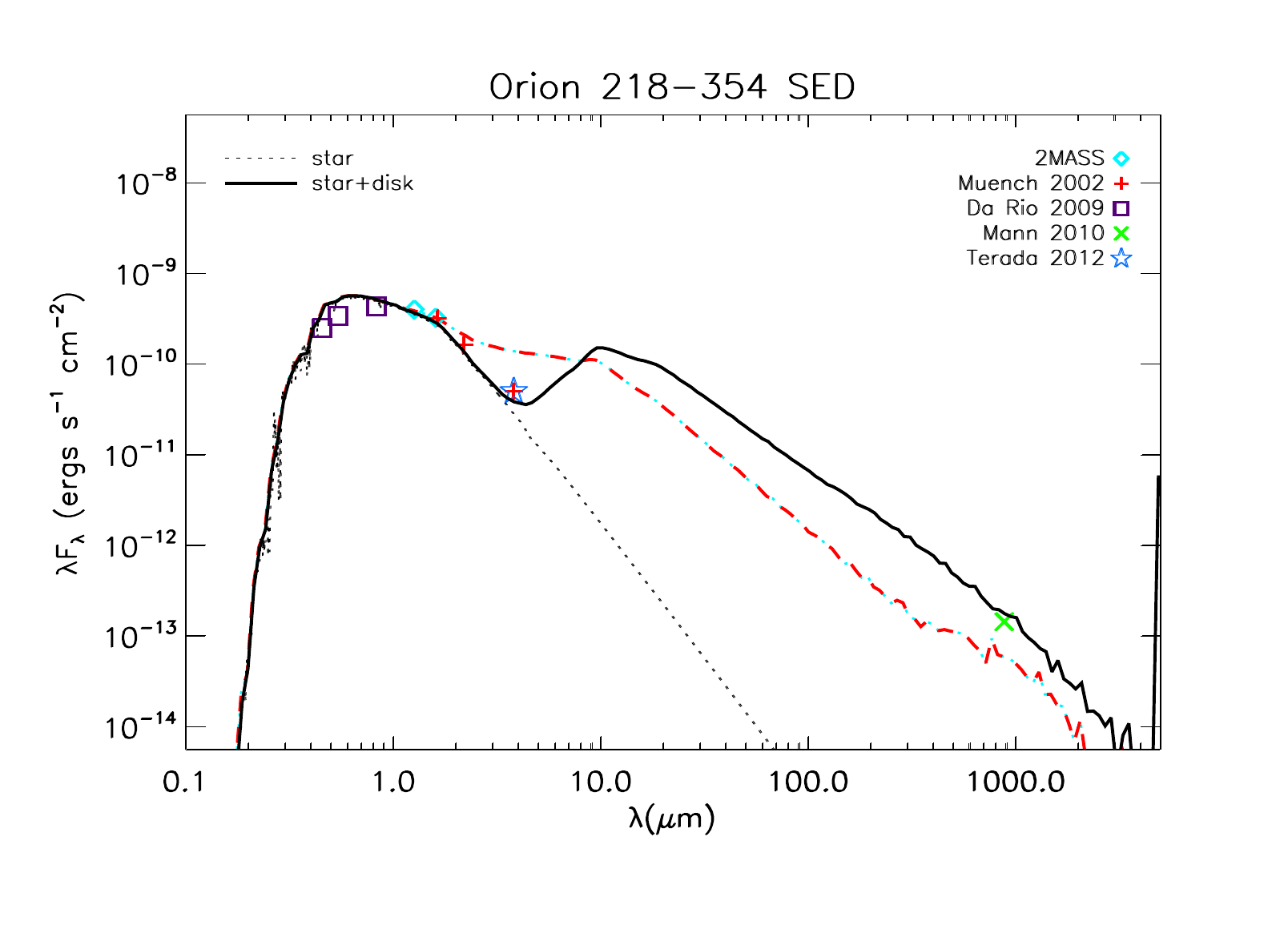}
\caption{Multiwavelength photometry of Orion 218-354 \citep{Muench:2002,Da-Rio:2009,Mann:2010,Terada:2012} overplotted on our best-fitting Whitney model output. The stellar spectrum (dashed black line) and best-fit model (solid black line, Table 1) are shown. The dashed red line corresponds to the same disk model with r$_{disk, in}$=r$_{sub}$. The poor fit suggests that Orion 218-354 contains an inner $\sim$2AU clearing. \label{fig4}}
\end{figure}

\clearpage

\bibliographystyle{apj}

\end{document}